\documentclass[pra,twocolumn,amsmath,amssymb,superscriptaddress,longbibliography]{revtex4-1}
\usepackage{graphicx}
\usepackage{amsthm}
\usepackage{amsfonts}
\usepackage{amssymb}
\usepackage{array}
\usepackage{amsmath}
\usepackage{upgreek,relsize,rotating}
\usepackage{verbatim} 
\usepackage{hyperref}
\usepackage{color}
\usepackage{bbold}
\usepackage{epstopdf}
\usepackage{mathtools}
\usepackage{booktabs}
\usepackage{bm,mathptmx,enumitem}
\usepackage[utf8]{inputenc}

\newcommand{\ket}[1]{\left|{#1}\right>}
\newcommand{\bra}[1]{\left<{#1}\right|}

\newcommand{\opinner}[3]{\left<{#1}\vphantom{#1}\vphantom{#3}\right|{#2}\left|{#3}\vphantom{#1}\vphantom{#3}\right>}
\newcommand{\tr}[1]{\mathrm{tr}\{#1\}}

\newcommand{\us}{\textscale{1.5}{\textunderscore}}

\def\bra#1{\langle #1|}
\def\ket#1{|#1 \rangle}

\begin{document}
\title{Objective Compressive Quantum Process Tomography}
\author{Y.~S.~Teo}
\email{ys\_teo@snu.ac.kr}
\affiliation{Department of Physics and Astronomy, Seoul National University, 08826 Seoul, Korea}
\author{G.~I.~Struchalin}
\email{struchalin.gleb@physics.msu.ru}
\affiliation{Quantum Technologies Centre, and Faculty of Physics, Moscow State University, 119991, Moscow, Russia}
\author{E.~V.~Kovlakov}
\affiliation{Quantum Technologies Centre, and Faculty of Physics, Moscow State University, 119991, Moscow, Russia}
\author{D.~Ahn}
\affiliation{Department of Physics and Astronomy, Seoul National University, 08826 Seoul, Korea}
\author{H.~Jeong}
\affiliation{Department of Physics and Astronomy, Seoul National University, 08826 Seoul, Korea}
\author{S.~S.~Straupe}
\affiliation{Quantum Technologies Centre, and Faculty of Physics, Moscow State University, 119991, Moscow, Russia}
\author{S.~P.~Kulik}
\affiliation{Quantum Technologies Centre, and Faculty of Physics, Moscow State University, 119991, Moscow, Russia}
\author{G.~Leuchs}
\affiliation{Max-Planck-Institut f\"ur die Physik des Lichts, Staudtstra\ss e 2, 91058 Erlangen, Germany}
\affiliation{Institute of Applied Physics of the Russian Academy of Sciences, 603950, Nizhny Novgorod, Russia}
\author{L.~L.~S{\'a}nchez-Soto}
\affiliation{Max-Planck-Institut f\"ur die Physik des Lichts, Staudtstra\ss e 2, 91058 Erlangen, Germany}
\affiliation{Departamento de \'Optica, Facultad de F\'{\i}sica,	Universidad Complutense, 28040 Madrid, Spain}
\date{\today}

\begin{abstract}
We present a compressive quantum process tomography scheme that fully characterizes any rank-deficient completely-positive process with no \emph{a priori} information about the process apart from the dimension of the system on which the process acts. It uses randomly-chosen input states and adaptive output von Neumann measurements. Both entangled and tensor-product configurations are flexibly employable in our scheme, the latter which naturally makes it especially compatible with many-body quantum computing. Two main features of this scheme are the certification protocol that verifies whether the accumulated data uniquely characterize the quantum process, and a compressive reconstruction method for the output states. We emulate multipartite scenarios with high-order electromagnetic transverse modes and optical fibers to positively demonstrate that, in terms of measurement resources, our assumption-free compressive strategy can reconstruct quantum processes almost equally efficiently using all types of input states and basis measurement operations, operations, independent of whether or not they are factorizable into tensor-product states.
\end{abstract}
\maketitle

{\it Introduction.---}Quantum computers and devices~\cite{Ladd:2010aa,Campbell:2017aa,Ladd:2010aa,Lekitsch:2017aa} employ a series of logic gate operations~\cite{Schafer:2018aa,Shi:2018aa,Ono:2017aa,Patel:2016aa,Fiurasek:2008fg} to carry out rapid computations using $d$-dimensional many-body systems, such as qubit ensembles. The quality of computed results hinges on the reliability of tomographic certifications of these gates, each of which is a quantum process $\Phi$ represented by a positive Choi--Jamio{\l}kowski operator $\rho_\Phi$~\cite{Choi:1975aa,Jamiolkowski:1972aa}, naively requiring $O(d^4)$ informationally complete (IC) measurements~\cite{Chuang:2000fk,OBrien:2004aa,Fiurasek:2001dn,Poyatos:1997aa} that are too resource-intensive to be implemented for large $d$. Ancilla-~\cite{Altepeter:2003aa,Leung:2003aa,D'Ariano:2003aa,D'Ariano:2001aa} and error-correction-based~\cite{Omkar:2015aa,Omkar:2015qc,Mohseni:2007aa,Mohseni:2006aa} quantum process tomography (QPT) were introduced to circumvent this problem. For highly-specific property prediction tasks, probing selected elements of $\rho_\Phi$ is another option~\cite{Kim:2018sw,Gaikwad:2018aa,Bendersky:2013aa,Schmiegelow:2011aa,Bendersky:2009aa,Bendersky:2008aa}. One attempt to directly reduce the measurement cost of QPT with non-IC measurements and entropy methods was reported~\cite{Teo2011aa}. Simultaneously, the method of compressed sensing~\cite{Donoho:2006cs,Candes:2006cs,Candes:2009cs,Gross:2010cs,Kalev:2015aa,Steffens:2017cs,Riofrio:2017cs} was applied to QPT~\cite{Baldwin:2014aa,Rodionov:2014aa,Shabani:2011aa} to reconstruct low-rank or sparse quantum processes with a small set of specialized compressive measurements. However, this concept only works under the assumption that $\rho_\Phi$ should either possess a rank no larger than some known integer $r$, or be sparse in some known basis of known sparsity, all of which demand reliable evidence. So, target guesses are needed to validate all reconstructions since there existed no self-consistent verification protocol.

In this Letter, we implement a compressive state-reconstruction-assisted quantum process tomography (ACTQPT)~scheme that requires no \emph{a priori} rank or sparsity knowledge, or any other precarious assumptions about $\Phi$ for that matter. The standard ancilla-free framework shall be considered here, which consists of input states $(\rho_\textsc{in})$ and output-state von~Neumann basis measurements that can be feasibly implemented in practice. Our scheme comprises an adaptive compressive (state) tomography~(ACT) protocol~\cite{Ahn:2019aa,Ahn:2019ns} that reconstructs the unknown output states $(\rho_\textsc{out}=\Phi[\rho_\textsc{in}])$ from adaptively chosen bases, and an informational completeness certification that determines whether the process estimator $\widehat{\rho}_\Phi$ (distinguished with a hat from the true process operator $\rho_\Phi$) is uniquely characterized by the accumulated dataset or not. This can be achieved with semidefinite programs~\cite{Vandenberghe:1996sd,Boyd:2004qd}. If $\widehat{\rho}_\Phi$ is not unique, the scheme is repeated with different linearly independent input states until $\widehat{\rho}_\Phi$ is unique. We also develop a product ACTQPT scheme~(PACTQPT) that adopts product input states and basis measurements suitable for realistic many-body implementations that avoid sophisticated entanglement manipulation and control. We run (P)ACTQPT in an emulated many-body setting using Hermite-Gaussian transverse degrees of freedom, which are high-order electromagnetic modes capable of experimentally enacting multiqubit processes with optical fibers. For all tested processes, both ACTQPT and PACTQPT perform comparably and are highly compressive relative to $O(d^4)$.

\begin{figure*}[t]
	\centering
	\includegraphics[width=1.6\columnwidth]{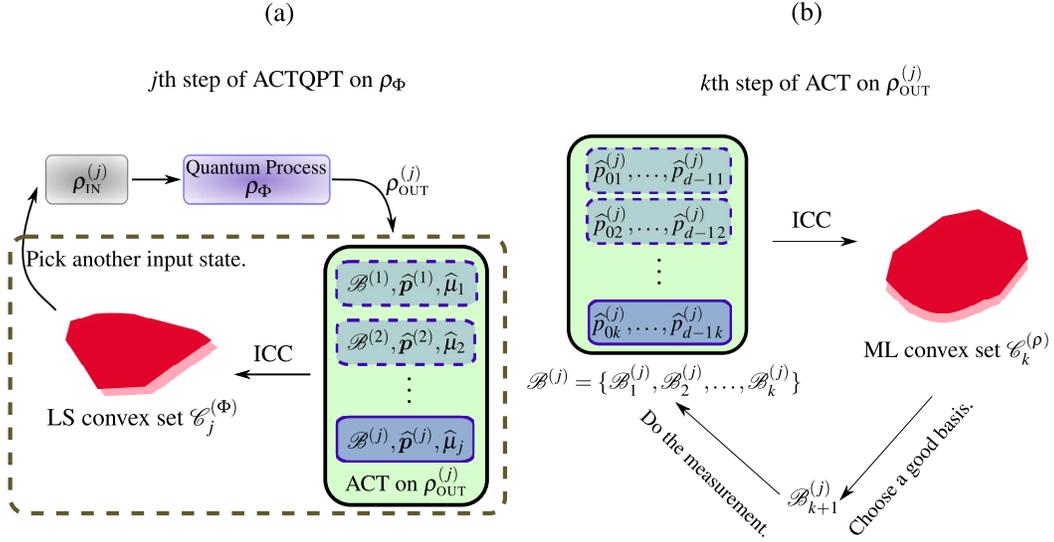}
	\caption{(a)~A flowchart of ACTQPT. An input state $\rho_\textsc{in}$ through the unknown $\Phi$ leads to an unknown output state $\rho_\textsc{out}$ that is compressively characterized with ACT at every step. The accumulated measurement bases sequences $\{\mathcal{B}^{(j')}\}$, ML probabilities $\{\widehat{\bm{p}}^{(j')}\}$ and estimated $\{\widehat{\mu}_{j'}\}$ are analyzed in ICC to deterministically decide if the associated LS convex set $\mathcal{C}^{(\Phi)}$ is singleton or not. If not, ACTQPT picks a new input state for more data collection. ACT is a sub-routine run for each $\rho_\textsc{out}^{(j)}$, which itself requires iterations. A schematic of the $k$th ACT iteration is shown in (b). It first performs ICC with the accumulated basis dataset to determine whether or not the ML convex set $\mathcal{C}^{(\rho)}$ is singleton. If not, the next basis measurement is chosen as the eigenbasis of a minENT estimator.}
	\label{fig:acqpt}
\end{figure*}

{\it Compressive state-reconstruction-assisted quantum process tomography.---}ACTQPT is an iterative scheme that completely and unambiguously characterizes any unknown rank-$r$ $\rho_\Phi$ based solely on data acquired from measuring output states $\{\rho^{(j')}_\textsc{out}\}$ as a consequence of the unknown process $\Phi$, and nothing else. For this purpose, the scheme is armed with two important components (see Fig.~\ref{fig:acqpt}). The first component is the ACT scheme~\cite{Ahn:2019aa,Ahn:2019ns} that chooses a compressive sequence of optimal von Neumann measurements to efficiently characterize every output state. In every step, it first certifies if the accumulated data uniquely characterize, say, $\rho_\textsc{out}$ after feeding $\rho_\textsc{in}$ to $\Phi$. This \emph{informational completeness certification}~(ICC) analyzes the convex space $\mathcal{C}^{(\rho)}$ of quantum states that give the same maximum-likelihood~(ML)~\cite{Banaszek:1999ml,Rehacek:2007ml,Teo:2011me,Teo:2015qs,Shang:2017sf} probabilities according to the accumulated dataset in order to compute an indicator $s^{(\rho)}_\textsc{cvx}$. If $s^{(\rho)}_\textsc{cvx}=0$, then $\mathcal{C}^{(\rho)}$ is a singleton---a set containing one operator---and ACT terminates. Otherwise the next optimal measurement is chosen as the eigenbasis of the minimum-von-Neumann-entropy~(minENT) estimator over $\mathcal{C}^{(\rho)}$. This generally leads to very efficient ACT convergences~\cite{Ahn:2019ns,Huang:2016aa,Tran:2016aa}.

Open-system quantum processes with losses are usually not trace-preserving~(TP), so that $\tr{\rho_\textsc{out}}\leq1$. After analyzing $j$ output states, their relative trace values $\{\mu_{j'}=\tr{\rho^{(j')}_\textsc{out}}\}^j_{j'=1}$ can be estimated (up to a common multiplicative factor), for example, by normalizing the total measured counts for all the different output states under a fixed measurement duration provided the count rate is sufficiently high. We now discuss the second crucial component, which is another ICC over the convex space $\mathcal{C}^{(\Phi)}$ of quantum processes that are consistent with the set of ML basis probabilities $\{\widehat{\bm{p}}^{(j')}\}$ and estimated $\{\widehat{\mu}_{j'}\}$ derived from all ACT runs so far. For the $j$th $\rho_\textsc{in}$, the operators in $\mathcal{C}^{(\Phi)}_j$ are those that give the same least-square (LS) value
\begin{equation}
\mathcal{D}_\text{min}=\min_{\widehat{\Phi}}\left\{\sum^j_{j'=1}\sum^{K_{j'}}_{k'=1}\sum^{d-1}_{l'=0}\left(\langle b^{(j')}_{l'k'}|\widehat{\Phi}[\rho^{(j')}_\textsc{in}]|b^{(j')}_{l'k'}\rangle-\widehat{\mu}_{j'}\widehat{p}^{(j')}_{l'k'}\right)^2\right\}\,,
\label{eq:LS}
\end{equation}
where all $K_j$ IC ACT measured bases in the set $\mathcal{B}^{(j)}=\{\mathcal{B}^{(j)}_1,\ldots,\mathcal{B}^{(j)}_{K_j}\}$ form a sequence of optimally chosen bases $\mathcal{B}^{(j)}_k=\{|b^{(j)}_{l'k}\rangle\langle b^{(j)}_{l'k}|\}^{d-1}_{l'=0}$ for $1\leq k\leq K_j$. The answer to ICC is a uniqueness indicator $s^{(\Phi)}_\textsc{cvx}$, with which we may conclude that $\widehat{\rho}_\Phi$ is unique if $s^{(\Phi)}_\textsc{cvx}=0$ and only then. Otherwise, ACTQPT picks another input state randomly, and the procedures of ACT and ICC over $\mathcal{C}^{(\Phi)}$ are repeated until $s^{(\Phi)}_\textsc{cvx}=0$.

The validity of ICC that systematically certifies reconstruction uniqueness with data relies on the fact that both $\mathcal{C}^{(\Phi)}$ and $\mathcal{C}^{(\rho)}$ are convex sets. We present a brief argument that guarantees this for a generic convex set $\mathcal{C}^{(X)}$ of bounded operators $X$, which may refer to either $\rho$ or $\rho_\Phi$ in our context. One can define a simple linear function $f=\tr{XZ}$, where $Z$ is another (positive) operator. Under this condition, it is known that the minimizing and maximizing $f$ over $\mathcal{C}^{(X)}$ give unique boundary answers, which we hereby denote by $f_\text{min}$ and $f_\text{max}$ respectively. It remains to show that if $s_\textsc{cvx}\propto f_\text{max}-f_\text{min}$, then (i)~$s_\textsc{cvx}$ does not increase with decreasing volume size $s$ of $\mathcal{C}^{(X)}$ for noiseless data, and (ii)~$s_\textsc{cvx}=0\leftrightarrow s=0$ (singleton condition) with any data. For (i), we observe that as noiseless data accumulate, and hence more distinct linear physical-probability constraints are imposed on $X$, $f_{\text{max}}$ decreases and $f_{\text{min}}$ increases progressively owing to the linearity of $f$. It is clear that $s_\textsc{cvx}$ also decreases with $s$. Property~(ii) follows immediately by noticing that as long as $\mathcal{C}^{(X)}$ is convex and $f$ has no ill-behaved plateau structures (guaranteed by a randomly-chosen full-rank $Z$), then $s=0$ necessarily implies a singleton $\mathcal{C}^{(X)}$. Whether this singleton set contains the true $X$ or another operator close to it depends on whether noise is present in the data. Optimizing $f$ over $\mathcal{C}$ can readily be translated into a semidefinite program~(SDP)~\cite{Vandenberghe:1996sd,Boyd:2004qd}.

ACTQPT is the whole iterative package \{entangled $\rho_\textsc{in}$s, ACT, ICC over $\mathcal{C}^{(\Phi)}$\}. On the other hand, the compressive measurement sequences obtained from ACT typically constitute entangled bases and are difficult to implement without sophisticated global entangling operations. Thus, a much more attractive alternative is to enforce a tensor-product local structure on all measurement bases, which turns ACT into its product counterpart~(PACT) that is experiment-friendly. This gives rise to another scheme that is much more suitable for many-body systems and quantum devices, namely the product version of ACTQPT~(PACTQPT) that requires only subsystem manipulations: \{product $\rho_\textsc{in}$s, PACT, ICC over $\mathcal{C}^{(\Phi)}$\}.

\begin{figure}[t]
	\centering
	\includegraphics[width=0.9\columnwidth]{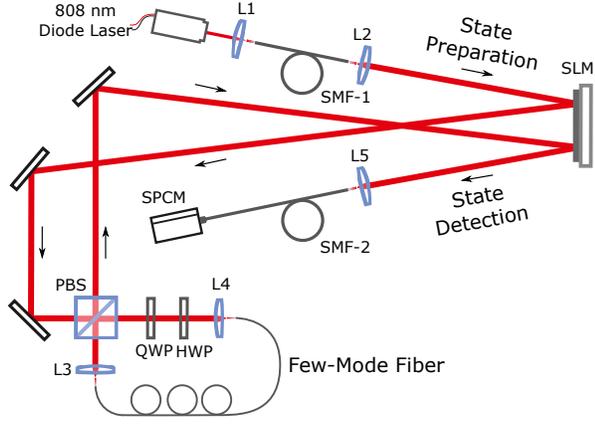}
	\caption{\label{fig:expt}Experimental setup for preparation and measurement of spatial-qudit states for rank-deficient quantum processes with a few-mode fiber. State preparation was realized with a hologram displayed on the top half of the SLM. A subsequent state detection was performed by the hologram on the SLM's bottom half followed by a single-mode fiber (SMF-2) and a single-photon counting module (SPCM).}
\end{figure}

{\it Numerical procedures.---}We shall explicitly state the operational iterative protocol of (P)ACTQPT pictorialized in Fig.~\ref{fig:acqpt}. At the $j$th step, (1)~all measured bases $\{\mathcal{B}^{(1)},\ldots,\mathcal{B}^{(j)}\}$, ML probabilities $\{\widehat{\bm{p}}^{(1)},\ldots,\widehat{\bm{p}}^{(j)}\}$ and estimates of relative trace values $\{\widehat{\mu}_1,\ldots,\widehat{\mu}_j\}$ of the output states, all obtained from (P)ACT, are first used to compute the LS value $\mathcal{D}_\text{min}$ defined in Eq.~\eqref{eq:LS} over the space of quantum processes. (2)~The resulting LS (non-TP) probabilities $\{\widehat{\bm{p}}^{(1)}_\textsc{ls},\ldots,\widehat{\bm{p}}^{(j)}_\textsc{ls}\}$ that correspond to $\mathcal{D}_\text{min}$, together with the measured bases, are next subject to ICC, which are two SDPs that lead to $s^{(\Phi)}_{\textsc{cvx},j}$ by minimizing and maximizing $f=\tr{\rho'_\Phi Z}$ for some fixed $d^2$-dimensional full-rank state $Z$ over the LS convex set $\mathcal{C}^{(\Phi)}_j$ that is defined by (a)~the complete-positivity quantum constraint $\rho'_\Phi\geq0$, (b)~the linear LS constraints $\langle b^{(j')}_{l'k'}|\widehat{\Phi}[\rho^{(j')}_\textsc{in}]|b^{(j')}_{l'k'}\rangle=\widehat{p}^{(j')}_{\textsc{ls},l'k'}$ for $0\leq l'\leq d-1$, $1\leq k' \leq K_{j'}$ and $1\leq j'\leq j$, and (c)~the trace constraint $\tr{\rho'_\Phi}=\tr{\widehat{\rho}^{(\textsc{ls})}_\Phi}$ for a particular LS process estimator $\widehat{\rho}^{(\textsc{ls})}_\Phi\in\mathcal{C}^{(\Phi)}_j$, where this final constraint mainly serves to stabilize the incorporation of the LS constraints. (3) (P)ACTQPT terminates when $s^{(\Phi)}_{\textsc{cvx},j}$ is less than some threshold value. Otherwise it proceeds to carry out (P)ACT for a new randomly-chosen input state, and $j$ is raised by one. 

The (P)ACT subprotocol is yet another self-consistent scheme that reconstructs any quantum state through adaptive compression. After $\rho^{(j)}_\textsc{in}$ is evolved by the unknown $\Phi$, the corresponding unknown $\rho^{(j)}_\textsc{out}$ undergoes an adaptive compressive iterative run. Starting with $k=1$, (1) the scheme measures the basis $\mathcal{B}_k$ and collects the relative frequency data $\sum^{d-1}_{l'=0}\nu^{(j)}_{l'k}=1$. (2)~From the accumulated dataset $\{\nu^{(j)}_{0k'},\ldots,\nu^{(j)}_{d-1\,\,k'}\}^k_{k'=1}$, the corresponding physical ML probabilities $\{\widehat{p}^{(j)}_{0k'},\ldots,\widehat{p}^{(j)}_{d-1\,\,k'}\}^k_{k'=1}$ are computed. (3)~These are then used to perform ICC, where the uniqueness indicator $s^{(\rho)}_{\textsc{cvx},k}$ is obtained by running two SDPs that find the unique minimum and maximum values of $f=\tr{\rho' Y}$, with a fixed $d$-dimensional full-rank state $Y$, over the ML convex set $\mathcal{C}^{(\rho)}_k$, which is defined by (a)~the positivity quantum constraint $\rho'\geq0$, (b) unit-trace constraint $\tr{\rho'}=1$, and (c)~the linear ML constraints $\langle b^{(j)}_{l'k'}|\rho'|b^{(j)}_{l'k'}\rangle=\widehat{p}^{(j)}_{l'k'}$ for $0\leq l'\leq d-1$ and $1\leq k'\leq k$. (4)~(P)ACT terminates if $s_{\textsc{cvx},k}$ is less than some prechosen threshold. Otherwise, it proceeds to choose the next optimal basis $\mathcal{B}_{k+1}$ to measure in the $(k+1)$th step. (5)~This is the adaptive stage, which first finds the minENT estimator over $\mathcal{C}^{(\rho)}_k$. From here, ACT directly takes $\mathcal{B}^{(j)}_{k+1}$ to be the eigenbasis of the minENT estimator, whereas PACT defines $\mathcal{B}^{(j)}_{k+1}$ as the nearest tensor-product basis to this optimal eigenbasis according to some distance metric of choice. After which, $k$ is raised by one and (P)ACT repeats itself until $s^{(\rho)}_{\textsc{cvx},k}$ is sufficiently small. Appendices~\ref{app:ML} and \ref{app:minENT} reveal more details on the ML and minENT procedures.

{\it Numerical and experimental results and analysis.---}The figure of merit for analyzing the performances of (P)ACTQPT is the IC number of measurement configurations $(M_\textsc{ic})$ needed to unambiguously reconstruct $\Phi$, which is the grand total of output-state measurement outcomes in a full (P)ACTQPT run. We experimentally tested (P)ACTQPT using transverse Hermite-Gaussian~(HG) spatial degrees of freedom to flexibly emulate complex quantum systems of various dimensions $d$. The corresponding continuous Hilbert space is discretized with the basis of these transverse modes. Figure~\ref{fig:expt} shows the schematic experimental setup. Attenuated light from a diode laser of 808nm wavelength was filtered by a single-mode fiber (SMF-1) and then collimated by an aspheric lens L2. The top half of the SLM (Holoeye Pluto) generates the desired spatial state of the photon and the bottom half followed by a single-mode fiber (SMF-2) implements projective measurements of the transformed state as in~\cite{Bent:2015experimental, Macarone:2019experimental}.

\begin{figure}[t]
	\centering
	\includegraphics[width=1\columnwidth]{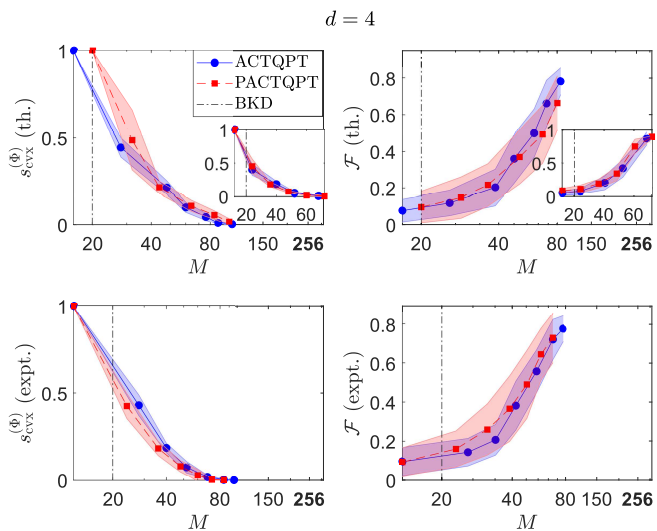}
	\caption{\label{fig:d4}Numerical and experimental plots and 1-$\sigma$ error regions of both $s^{(\Phi)}_\textsc{cvx}$ and $\mathcal{F}$ for a rank-4 two-qubit process $(d=4)$ against $M$, averaged over 40 and 20 runs respectively. Numerical results with noiseless data are shown in the insets, whereas simulations with noisy data~($\eta=0.01$) explain the actual experiments very well. The results are compared with the optimal benchmark $(M_\textsc{bkd}=20)$. The $s_\textsc{cvx}$ termination threshold is set to $10^{-3}$.%(4 positive eigenvalues: 0.8721, 0.1062, 0.0160, 0.0057)
	}
\end{figure}

\begin{figure}[t]
	\centering
	\includegraphics[width=1\columnwidth]{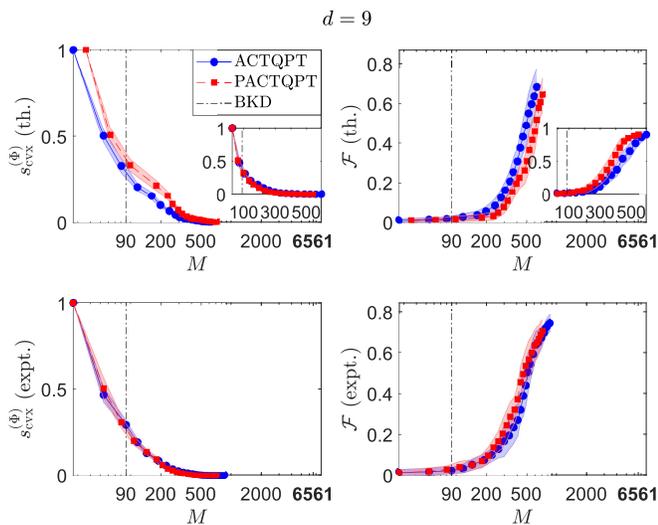}
	\caption{\label{fig:d9}Numerical and experimental plots and 1-$\sigma$ error regions of both $s^{(\Phi)}_\textsc{cvx}$ and $\mathcal{F}$ for a rank-10 two-qutrit process $(d=9)$ against $M$. The optimal BKD benchmark here is $M_\textsc{bkd}=90$, and the average $M_\textsc{ic}$s for both ACTQPT and PACTQPT are again comparable and about an order of magnitude smaller than $d^4=6561$. All other figure specifications are otherwise the same as those of Fig.~\ref{fig:d4}.%(10 positive eigenvalues: 0.9253, 0.0252, 0.0200, 0.0125, 0.0082, 0.0043, 0.0025, 0.0014, 0.0005, 0.0002)
	}
\end{figure}

In Figs.~\ref{fig:d4} and \ref{fig:d9}, we compare (P)ACTQPT with a recently proposed optimal benchmark provided by the Baldwin-Kalev-Deutsch~(BKD) scheme~\cite{Baldwin:2014aa}, requiring both the unitarity assumption and $M_\textsc{bkd}=d^2+d$ non-projective entangling measurements. Few-mode optical fibers operating at detuned photonic wavelengths, which behave as rank-deficient processes acting on two-qubit ($d=4$) and two-qutrit systems ($d=9$), are characterized with (P)ACTQPT. To realize these processes, we coupled the photons after the PBS into a meter-long few-mode optical fiber. The utilized fiber~(SMF-28) supported the propagation of HG modes of four lowest orders at the 808nm operating wavelength, which is detuned from the designed wavelength of 1.5$\mu$m for the single-mode regime. We were able to control the output mode content by altering input polarization using the half-wave~(HWP) and quarter-wave~(QWP) plates, followed by the PBS. This is possible due to cross-coupling between polarization and spatial degrees of freedom in optical fibers~\cite{Carpenter:2013mode}. Noisy theoretical simulations are performed to model the observed experimental results, where statistical noise on each output-state von Neumann basis projector $\ket{b^{(j')}_{l'k'}}\bra{b^{(j')}_{l'k'}}$ is modeled as an independent Gaussian distribution of a fixed effective standard deviation $\eta$ (see Appendix~\ref{app:sim} for more technical details). Noiseless simulation results of rank-1 processes, each taken as the largest-eigenvalue projector of the actual rank-deficient $\rho_\Phi$ determined \emph{via} overcomplete tomography, are presented for idealized comparisons. The fidelity $\mathcal{F}=\tr{\sqrt{\widehat{\rho}_\Phi}\rho_\Phi\sqrt{\widehat{\rho}_\Phi}}/(\tr{\widehat{\rho}_\Phi}\tr{\rho_\Phi})$ is defined between two general non-TP process operators. The average $M_\textsc{ic}$ values of both ACTQPT and PACTQPT for Figs.~\ref{fig:d4} and \ref{fig:d9} are presented in Tab.~\ref{tab:MIC}.
\begin{table}[t]
\begin{tabular}{rlccc}
	&& {Th. $(\eta=0)$} & {Th.} $(\eta=0.01)$ & {Expt.}\\
	\hline\\
	&{ACTQPT}  & $88.8\pm9.9$ & $126.0\pm15.6$ & $114.8\pm11.9$\\
	\begin{rotate}{90}\!\!\!\!($d=4$)\end{rotate}&{PACTQPT} & $97.4\pm12.3$ & $133.3\pm20.9$ & $122.8\pm16.2$\\[5ex]
	&& {Th. $(\eta=0)$} & {Th.} $(\eta=0.005)$ & {Expt.}\\
	\hline\\
	&{ACTQPT}  & $621.4\pm56.3$ & $884.7\pm108.9$ & $927.0\pm24.7$\\
	\begin{rotate}{90}\!\!\!\!($d=9$)\end{rotate}&{PACTQPT} & $613.3\pm37.4$ & $864.0\pm99.9$ & $808.8\pm46.3$
\end{tabular}
\caption{\label{tab:MIC}Table of average $M_\textsc{ic}$ values and their statistics for the two-qubit and two-qutrit fiber processes studied in Figs.~\ref{fig:d4} and \ref{fig:d9}.}
\end{table}

We also found that there is actually not a very big difference in the average $M_\textsc{ic}$ relative to $O(d^4)$ between ACTQPT and PACTQPT for the tested processes. Furthermore, different choices of optimal adaptive bases and confounding experimental processes can result in varying relative performances. We attribute their performance similarity to the inherent input-output characteristic framework of these QPT schemes, which can limit further enhancements with entangled input states and basis measurements. Our results also demonstrate that (P)ACTQPT is highly compressive. Both the two-qubit and two-qutrit experiments respectively show over 50\% and 85\% reduction in IC measurement resources relative to $O(d^4)$ without spurious prior assumptions \emph{of any sort} about the processes. They also indicate that if one takes the rank of an unknown process for granted, one presumably gets away with a further $\approx90\%$ reduction in measurement resources that are, especially in our case, of paramount importance for validating this rank assumption and final reconstruction answer.

{\it Conclusion.---}We have successfully demonstrated a compressive state-reconstruction-assisted quantum process tomography method that requires no \emph{a priori} assumptions about the unknown quantum process, in particular not the frequently taken-for-granted rank/sparsity assumption, to fully characterize the process with much fewer measurement resources than the fourth power of the system dimension. It involves a uniqueness reconstruction certification procedure over the general completely-positive quantum-process space and a recently established adaptive compressive state tomography scheme. Our results with experimentally implemented processes indicate that the compressive method works equally well with both entangled and product input states and output measurement resources even in the presence of noise. This allows one to implement this method in practice using uncorrelated quantum resources without precise and expensive global entangling operations.

\begin{acknowledgments}
	We acknowledge financial support from the BK21 Plus Program (21A20131111123) funded by the Ministry of Education (MOE, Korea) and National Research Foundation of Korea (NRF), the NRF grant funded by the Korea government (MSIP) (Grant No. NRF-2019R1H1A3079890), RFBR (project No. 19-32-80043), Mega-grant of the Ministry of Education and Science of the Russian Federation (Contract No.14.W03.31.0032), and the Spanish MINECO (Grant FIS2015-67963-P and PGC2018-099183-B-I00).
\end{acknowledgments}

\appendix

\section{Maximum-likelihood physical probabilities}
\label{app:ML}

The normalized frequency data $\mathbb{D}=\{\nu^{(j')}_{l'k'}\}$ collected in ACT are noisy and, almost always, do not correspond to any physical quantum state. We therefore need to estimate the correct physical probabilities $\widehat{p}^{(j')}_{l'k'}$ that asymptotically approach the true ones in the limit of large sampling copies.

A good statistical approach for dealing with a finite number of sampling events is to first identify the likelihood function $L(\mathbb{D}|\rho')$ for the experiment, which may be taken as the multinomial form in the situation where $N$ is a fixed number and all copy detection of a basis measurement is independent and identically distributed. The corresponding concave multinomial (normalized) log-likelihood expression reads
\begin{equation}
\log L(\mathbb{D}|\rho')=\sum_{j',l',k'}\nu^{(j')}_{l'k'}\log \opinner{b^{(j')}_{l'k'}}{\rho'}{b^{(j')}_{l'k'}}\,.
\label{eq:f}
\end{equation}
This likelihood function has the meaning of a conditional probability of obtaining $\mathbb{D}$ given the state $\rho'$, and maximizing this function over all quantum states shall then give the most-likely physical state $\widehat{\rho}\geq0$ that gives $\mathbb{D}$. The estimated ML physical probabilities are then $\widehat{p}^{(j')}_{l'k'}=\opinner{b^{(j')}_{l'k'}}{\widehat{\rho}}{b^{(j')}_{l'k'}}$.

One may adopt the steepest-ascent algorithm~\cite{Rehacek:2007ml,Teo:2011me,Teo:2015qs} to maximize $L$. The most efficient algorithm to date, however, can be derived according to the accelerated projected-gradient recipe, where at its core is an augmented rapid-converging iteration of a likelihood maximization over all unit-trace Hermitian operators followed by a projection onto the unit-trace positive-operator space~\cite{Shang:2017sf}. Without restating the entire code, we refer the Reader to the GitHub page (\url{https://github.com/qMLE/qMLE}) for ready-to-use MATLAB codes, complete with tutorials and examples.

\section{Minimum-entropy optimization in ACT}
\label{app:minENT}

For completeness, we shall reiterate minENT procedure for ACT (found in \cite{Ahn:2019aa,Ahn:2019ns}) in this section. We recall the fundamental fact that minimizing a concave function over convex spaces is generally not a convex problem. The consequence of which is a non-unique optimal solutions to choose from the optimization. Semidefinite programs are therefore incompatible with such a problem. Nevertheless, we construct an equivalent and simple barrier method to minimize the von Neumann entropy $S(\rho')=-\tr{\rho'\log\rho'}$ over the ML convex set $\mathcal{C}^{(\rho)}$.

Following \cite{Teo:2011me}, we first consider the Lagrange function $\mathcal{D}=-\lambda S+\log L$ that is a sum of $S$ and the log-likelihood $\log L$ weighted by a small positive parameter $\lambda\ll1$, where the $\rho'$ dependence is dropped from all functions for notational convenience. We also note that for any non-IC dataset $\mathbb{D}$, the corresponding $L$~(or $\log L$) possesses a plateau structure over the quantum domain $\mathcal{C}^{(\rho)}$. A perfectly accurate minENT state estimator that both minimizes $S$ and remains in $\mathcal{C}^{(\rho)}$, the subspace of quantum states that give the same maximal $L$ value, therefore corresponds to a $\lambda$ that is infinitesimal. As this is never feasible in practice, we approximate this situation with a small finite $\lambda$ such that both conditions are satisfied with a finite precision.

Numerically, we may again make use of the superfast accelerated projected gradient method using the gradient operator $\updelta\mathcal{D}/\updelta\rho'$ for the minENT procedure instead of $\updelta \log L/\updelta\rho'$ for the usual ML optimization considered in~\cite{Shang:2017sf}. For this purpose, we supply a simple instruction manual to modify and use the open-source MATLAB code file \texttt{qse\us apg.m} that is available on \url{https://github.com/qMLE/qMLE}. The three important variables are \texttt{fval\us varrho}, \texttt{fval\us new} and \texttt{gradient}, which stores the function values of $\mathcal{D}$ evaluated with the \texttt{varrho} and \texttt{rho\us new} variables, and the gradient operator
\begin{equation}
\dfrac{\updelta\mathcal{D}}{\updelta \rho'}=\lambda(1+\log\rho')+\sum_{j',l',k'}\,\ket{b^{(j')}_{l'k'}}\dfrac{\nu^{(j')}_{l'k'}}{ \opinner{b^{(j')}_{l'k'}}{\rho'}{b^{(j')}_{l'k'}}}\bra{b^{(j')}_{l'k'}}
\label{eq:df}
\end{equation} 
evaluated with \texttt{varrho}. In order to minimize $\mathcal{D}$ using \texttt{qse\us apg.m}, we may simply overwrite the existing functional expressions [namely \texttt{-f'.*log(probs\us\ldots)} and \texttt{-qmt(\ldots)}] for the three variables with those in Eq.~\eqref{eq:f} and \eqref{eq:df}. 

\section{Numerical simulations}
\label{app:sim}

The simulation results presented come in two flavors. For each $d$, the noiseless simulations, which serve as background benchmarks, are generated with a true Choi-Jamio{\l}kowski process operator $\rho_\Phi$ that is rank-one, which approximates the would-be process operator for ideal fibers. Prior to the experimental (P)CQPT runs, the optical fibers are subjected to an overcomplete state tomography and the final estimators $\widehat{\rho}_\Phi^{(\mathrm{overcomp.})}$ are used to define the ``true'' process operators for the numerical simulations. These estimators are found to have a fast-decaying eigenvalue spectrum. Explicitly, the $d=4$ process operator has four positive eigenvalues 0.8721, 0.1062, 0.0160, and 0.0057; while the $d=9$ process operator has 10 positive eigenvalues 0.9253, 0.0252, 0.0200, 0.0125, 0.0082, 0.0043, 0.0025, 0.0014, 0.0005, and 0.0002.

In all of the experiments, the count of each projector is measured one projector at a time by maintaining a particular measurement configuration for some fixed duration. Hence, for the noisy simulations, noise on each basis projector $\ket{b^{(j')}_{l'k'}}\bra{b^{(j')}_{l'k'}}$ is modeled with a single-variable Gaussian distribution for its normalized measured relative frequencies with respect to $\rho^{(j')}_\textsc{out}$, with mean set to $p^{(j')}_{l'k'}$ and variance set to an effective value $\eta^2$ that is common to all projectors. The value of $\eta>0$ is chosen to match the experimental results in order to explain the actual observed noise with an effective Gaussian noise model. These simulations assume the true processes $\rho_\Phi\equiv\widehat{\rho}_\Phi^{(\mathrm{overcomp.})}$.

For the noiseless simulations, we assume the ideal situation where perfect fibers are used. These implies that the resulting quantum-process operators are all rank-one. Since $\widehat{\rho}_\Phi^{(\mathrm{overcomp.})}$ has rapidly-decaying eigenvalues, we define $\rho_\Phi$ to be the rank-one support of the largest eigenvalue of $\widehat{\rho}_\Phi^{(\mathrm{overcomp.})}$.

\end{document}